\documentclass[twocolumn,superscriptaddress,showpacs,amsmath,amssymb,prl]{revtex4}
\usepackage{graphicx}

\newcommand{\apjl}{\mbox{\it Astrophys. J.}}
\newcommand{\aap}{\mbox{\it Astron. Astrophys.}}
\newcommand{\araa}{\mbox{\it Annu. Rev. Astron. Astrophys.}}

\newcommand{\jcap}{\mbox{\it J. Cosmol. Astropart. Phys.}}
\newcommand{\mnras}{\mbox{\it Mon. Not. R. Astron. Soc.}}

\newcommand{\physrep}{\mbox{\it Phys. Rep.}}

\newcommand{\eV}{\rm{\, eV }}

\newcommand{\MeV}{\rm{\, MeV }}

\newcommand{\llu}{\rm{\, erg \, s^{-1}}}
\newcommand{\beq}{\begin{equation}}
\newcommand{\eeq}{\end{equation}}
\newcommand{\ba}{\begin{array}}
\newcommand{\ea}{\end{array}}

\newcommand{\ee}{\epsilon_{e,0}}

\def\be{\begin{equation}}
\def\ee{\end{equation}}
 \def\lsim{\raisebox{-0.3ex}{\mbox{$\stackrel{<}{_\sim} \,$}}}

\begin{document}

\title{Constraining Sources of Ultra High Energy Cosmic Rays Using
  High Energy Observations with the Fermi Satellite}

\author {Asaf Pe'er and  Abraham Loeb}
\affiliation{Harvard-Smithsonian Center for Astrophysics, MS-51, 60
  Garden Street, Cambridge, MA 02138, USA
}

\begin{abstract}
  We analyze the conditions that enable acceleration of particles to
  ultra-high energies, $\sim 10^{20}$~eV (UHECRs). We show that broad
  band photon data recently provided by {\it WMAP}, {\it ISOCAM}, {\it Swift} and
  {\it Fermi} satellites, yield constraints on the ability of active
  galactic nuclei (AGN) to produce UHECRs. The high energy (MeV -- GeV)
  photons are produced by Compton scattering of the emitted low energy
  photons and the cosmic microwave background or extra-galactic
  background light. The ratio of the luminosities at high and
  low photon energies can therefore be used as a probe of the physical
  conditions in the acceleration site.  We find that existing data
  excludes core regions of nearby radio-loud AGN as possible acceleration
  sites of UHECR protons. However, we show that giant radio lobes are not
  excluded. We apply our method to Cen~A, and show that acceleration
  of protons to $\sim 10^{20}$~eV can only occur at distances $\gtrsim
  100$~kpc from the core.
\end{abstract}


\date{\today}

\maketitle


The origin of ultra-high energy cosmic rays (UHECRs), with energies
$\gtrsim 10^{18.5}$~eV, is still under debate.  Astrophysical sources
of UHECRs are limited by two requirements: a strong magnetic field is
needed to confine the accelerated cosmic rays, while the magnetic
field cannot be too strong in order to avoid excessive synchrotron
radiation and photo-meson energy losses \cite{Hillas84}. In addition,
energy losses by photo-meson production through scattering off the
cosmic microwave background (CMB) limit the distance of UHECR sources
to $\lesssim 100$ּ~ּMpc (the so-called GZK cutoff)
\cite{Greisen66,ZK66}.  Several possible sources of UHECRs that
fulfill these constraints are discussed in the literature. The leading
candidates are gamma-ray bursts (GRBs) \cite{MU95,W95,W04}, low
luminosity gamma-ray bursts and hypernovae \cite{Wang+07, Murase+08,
  Chakraborti+11, LW11}, and active galactic nuclei (AGNs)
\cite{BS87,PS92, RB93,Norman+95,BGG06, MO10, Nemmen+10}.

The possibility that AGNs are the main source of UHECRs is challenged
by the need for an unusually high photon luminosity.  The required
strong magnetic field, combined with the assumption that the energy
density in the photon field is at least comparable to the energy
density in the magnetic field, constrain the power output from the
source to be $ L > 10^{45} \Gamma^2 \beta^{-1} \llu$ \cite{Lovelace76,
  Norman+95, W04b}. Here, $\Gamma$ and $\beta c$ are the Lorentz
factor and characteristic velocity within the source. Only very few
AGNs within the GZK horizon fulfill this requirement. A detailed
analysis of those AGNs whose position was found to be correlated with
the arrival direction of UHECRs \cite{Abraham+08}, resulted in a
similar conclusion \cite{Zaw+09}.

The underlying assumption of the above mentioned constraint is that
the energy density in the photon field is comparable or higher than
the energy density in the magnetic field. It is indeed difficult to
obtain direct constraints on the value of the magnetic fields in AGNs
and test this assumption.

In this {\it paper}, we re-analyze the constraints on AGNs as UHECR
sources by combining low energy (radio-optical) and high energy ($\sim
\MeV $~band) data, enabled by recent {\it Swift} and {\it Fermi}
observatories. These data provide constraints on the magnetic field
strength, and avoid the need to adopt the equipartition assumption. In
addition, the broad band data itself places constraints on the allowed
region within the AGN that can produce UHECRs. The constraints are
based on observable quantities, and thus overcome the inherent
uncertainty in estimating the magnetic field strength.

\paragraph*{Basic requirements.} Assuming that UHECRs acceleration
results from electromagnetic processes within an expanding plasma,
the requirement that the accelerated particles are confined to the
acceleration region is equivalent to the requirement that the
acceleration time, $t_{\rm acc} = \eta E^{ob}/ (\Gamma Z e B
c)$\footnote{This result is relevant for first-type Fermi
  acceleration. For second type Fermi mechanism, $t_{\rm acc} \propto
  E^{ob} \log (E^{ob}) $, where the linear term corresponds to the Larmor
  radius, while the logarithmic term originates from the number of
  cycles a particle has to undergo during the acceleration process.},
is shorter than the dynamical time, $t_{dyn} = r/\Gamma \beta
c$. This condition implies 
\beq
B r \geq \frac {\eta E^{ob} \beta} {Z e} = 3.3 \times 10^{17} \beta
\left( \frac{\eta E^{ob}_{20}} {Z} \right) \; \rm{G \, cm}.
\label{eq:1}
\eeq
Here, $E^{ob}  $ is the observed energy of the particle, $Z e$ is its
charge, $B$ is the magnetic field and $r$ is the characteristic size
of the acceleration region. $\eta \geq 1$ is a
dimensionless factor, whose exact value is determined by the
uncertain details of the acceleration mechanism\footnote{For example,
  in the non-relativistic diffusive shock acceleration mechanism, this
  factor corresponds to $\eta = (20/3)\beta^{-2}$ in the Bohm limit
  for a plane parallel shock\cite{BE87}.}. Here and below, we
use the convention $Q = 10^X Q_X$ in cgs units.

A second condition is that the acceleration time is shorter than all
relevant energy loss time scales. Energetic particles can lose their
energy via synchrotron emission, on a time scale $t_{\rm cool, syn} =
(6 \pi m_p^4 c^3 \Gamma A^4)/(\sigma_T m_e^2 B^2 E^{ob} Z^4)$.  Here,
$\sigma_T$ is Thomson's cross section, $m_p$ and $m_e$ are the proton
and electron masses and $A m_p$ is the mass of the nucleon (for iron
nuclei, $A=56$ and $Z=26$). The requirement $t_{\rm acc} < t_{\rm
  cool, syn}$ results in
\beq
E^{ob} \leq 2 \times 10^{20} \; \Gamma\, \eta^{-1/2}\,  B^{-1/2}\, A^2
\, Z^{-3/2} \eV.  
\label{eq:t_loss}
\eeq
Thus, both protons and iron nuclei can be accelerated to the highest
observed energies, $\sim 10^{20}$~eV, provided that the strength of
the magnetic field at the acceleration site does not exceed $B \,
\lsim$~few - few tens G.

In addition to synchrotron energy losses, energetic particles can in
principle lose their energy by interacting with the ambient photon
field and with other nuclei.  Interaction with the photon field can
result in energy losses by Compton scattering (which is typically a
negligible energy loss channel for UHECRs), photopair production
(Bethe-Heitler process), photo-production of mesons (mainly pions),
and photodisintegration of heavy nuclei. Interaction with other nuclei
can lead to spallation of heavy nuclei. These processes were studied
in details for radio-loud AGNs in \cite{AHST08}, and for radio-quiet
AGNs in \cite{PMM09}. The results of both these works show that
$\sim 10^{20}$~eV particles will survive all energy losses, provided
that the acceleration site is located at $\sim$ few - few tens of
parsecs from the core.

Assuming that electrons are being accelerated at the same acceleration
site of the UHECRs, synchrotron emission is expected. Indeed, in many
AGNs the peak of the synchrotron emission is clearly identified at the
radio or infrared bands. The peak of the synchrotron energy flux can be
approximated by assuming that the energetic electrons have a
characteristic Lorentz factor $\gamma_e$, 
\beq
(\nu F_\nu)_{\rm peak, syn} = \frac {n_e V}{4 \pi d_L^2} \left(\frac{4}{3}\right) c
  \sigma_T \gamma_e^2 \left(\frac {B^2}{8 \pi}\right)  \mathcal{D}^2.
\label{eq:2}
\eeq
Here, $n_e$ is the number density of energetic electrons, $V$ is the
volume of the emitting region, $d_L$ is the luminosity distance,
$\sigma_T$ is the Thomson cross section and $\mathcal{D} = [\Gamma
(1-\beta \cos(\theta^{ob})]^{-1}$  is the Doppler factor for an
observing angle $\theta^{ob}$ \footnote{We assume here that the
  observer is inside the lightcone. Otherwise, a factor
  $\mathcal{D}^4$ should appear in equation \ref{eq:2}.}. For an observer within the
light cone, $\theta^{ob} < \max(\Gamma^{-1}, \theta_{jet})$,
$\mathcal{D} \simeq \Gamma$, where $\theta_{jet}$ is the physical jet
opening angle \footnote{The factor $\mathcal{D}^2$ results from our
  simplified assumption of monochromatic emission. For a power law spectra, a
  somewhat more complicated  dependence on $\mathcal{D}$ is obtained;
  see, e.g., \cite{BK79}. This, however, does not affect our conclusions.}.


Recent {\it Swift} and {\it Fermi} observations revealed a second peak
in the spectra of many AGNs, centered at $\sim \MeV$
\cite[e.g.,][]{Abdo+09, Abdo+10, Abdo+10c}. This second peak results
from inverse-Compton (IC) scattering of low energy photons by the
energetic electrons; a hadronic origin can be excluded by the lack of
a neutrino counterpart \cite{Abbasi+11}.  Let us denote the flux ratio
of the IC and the incoming photons peak energies by $\bar R \equiv
(\nu F_\nu)_{\rm peak, IC} /(\nu F_\nu)_{\rm peak, in}$. We do not
assume that the incoming photons are necessarily the synchrotron
photons, and we thus denote their frequency by $\nu_{\rm in}$, which
can generally be different from $\nu_{\rm syn}$. In an IC process, the
outgoing photon energy is $\nu_{\rm IC} = (4/3) \gamma_{e}^2
\nu_{in}$, and the outgoing monochromatic (number) flux ratio of IC to
synchrotron emission is $(F_\nu)_{\rm peak, IC} /(F_\nu)_{\rm peak,
  in} = \tau \simeq \Delta l n_{e} \sigma_T$. Approximating the volume
of the acceleration region as cylindrical, $V = \pi r^2 \Delta l$, and
substituting these results in equation (\ref{eq:2}), one obtains
\beq 
(\nu  F_\nu)_{\rm peak, syn} = \frac {1}{4 \pi d_L^2} \frac{c {\bar R} B^2
   r^2}{8} \mathcal{D}^2.
\label{eq:3}
\eeq
Combined with the constraint set in equation (\ref{eq:1}),
one finds
\beq
(\nu  F_\nu)_{\rm peak, syn} \geq \frac {1}{4 \pi d_L^2} \frac{c \bar
  R}{8} \left(\frac{\eta E^{ob}}{Z q}\right)^2 \beta^2  
\mathcal{D}^2 .
\label{eq:4}
\eeq

\begin{figure}
\includegraphics[width=\linewidth]{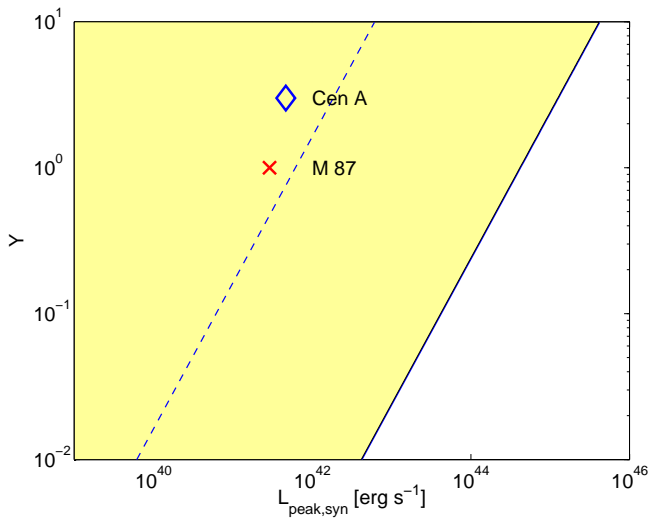}
\caption{Minimum peak luminosity of synchrotron emission in a source
  that is capable of accelerating UHECRs to $10^{20}$~eV, as a
  function of $Y = (\nu F_\nu)_{\rm peak, IC} /(\nu F_\nu)_{\rm peak,
    syn}$. The shaded (light yellow) region to the left of the solid
  line is excluded for $(\eta
  E_{20}^{ob}/Z) \beta \mathcal{D}= 1$. The dashed line is for iron nuclei
  (Z=26). The measured values for the FR-I radio galaxies Cen A and
  M87 are marked, based on data in Refs. \cite{Abdo+09,Abdo+10}.  }
\label{fig:1}
\end{figure}

\paragraph*{Emission from AGN cores.} A primary source of emission in
AGNs is the core, or inner parts of the jet. Due to the high
brightness temperature of the radio emission from this region, the
seed photons to IC scattering are the synchrotron photons. Thus, $\bar
R \equiv (\nu F_\nu)_{\rm peak, IC} /(\nu F_\nu)_{\rm peak, syn} = Y$,
where $Y = (4/3) \gamma_e^2 \tau$ is Compton parameter for an optical
depth $\tau$. Equation (\ref{eq:4}) can be written as a constraint on
the minimum synchrotron luminosity that is needed for a source to be
able to produce UHECRs,
\beq
\ba{lcl}
L_{\rm peak,syn} &\equiv & 4 \pi d_L^2 (\nu  F_\nu)_{\rm peak, syn}
\geq \frac{c Y}{8} \left(\frac{\eta E^{ob}}{Z q}\right)^2 \beta^2 
\mathcal{D}^2 \\
& = & 4.1 \times 10^{44} {Y} \left(\frac{\eta E_{20}^{ob}}{Z}\right)^2 \beta^2 
\mathcal{D}^2 \; \llu.
\ea
\label{eq:5}
\eeq
Figure \ref{fig:1} shows the allowed region of the synchrotron
luminosity as a function of $Y$ \footnote{The allowed region may be
  smaller, due to uncertainty that exists in the value of $\eta
  \beta \mathcal{D}$.}. Two nearby sources, Cen~A and M87,
have good enough spectral coverage for a reliable measurement of the
synchrotron peak and the $Y$ parameter. Both fall within the excluded
region for acceleration of UHECRs to $10^{20}$~eV, unless UHECRs are
composed of heavy nuclei. We point that for $Y \gg 1$, the
electrons can rapidly cool, leading to a broad band emission
spectrum \cite{PMR06}. In this scenario, it would not be possible to
identify both the synchrotron and the IC peaks, and the analysis could
not be carried out.

\paragraph*{Emission from AGN lobes.} While acceleration of UHECRs in
the inner regions of AGNs are excluded by the data on Cen~A and M87,
jets from radio loud AGNs extend out to hundreds of kpc. Thus,
additional possible acceleration sites exist in the turbulent outflow
or at the termination shock of the giant AGN lobes.

\begin{figure}
\includegraphics[width=\linewidth]{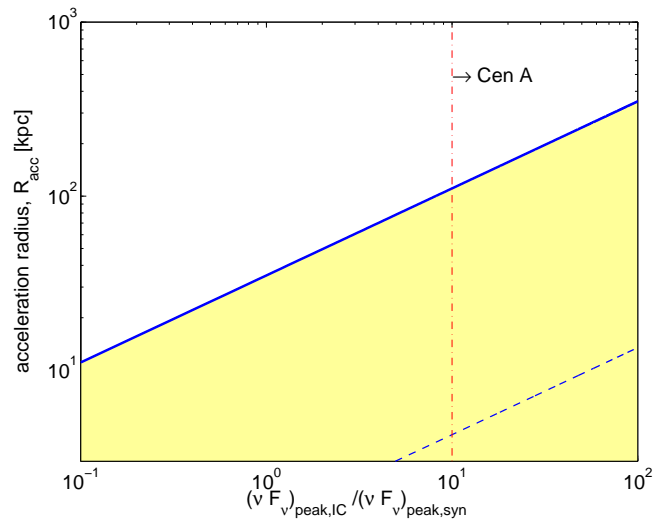}
\caption{Minimum radius of acceleration region that allow acceleration
  of UHECRs to $10^{20}$~eV, as a function of the ratio $(\nu
  F_\nu)_{\rm peak, IC} /(\nu F_\nu)_{\rm peak, syn}$. The shaded
  (light yellow) region below the solid line is excluded for $(\eta
  E_{20}^{ob}/Z) \beta \mathcal{D} = 1$. The dashed line is for iron
  nuclei (Z=26). The dash-dotted (vertical) line gives the minimum
  ratio of $(\nu F_\nu)_{\rm peak, IC} /(\nu F_\nu)_{\rm peak, syn}
  \geq 10$, measured for Cen A \cite{Abdo+10b}. Thus, Cen A can be the
  source of UHECRs, provided that the acceleration occurs beyond $\sim
  110$~kpc.  }
\label{fig:2}
\end{figure}

Due to the large distance from the source, the seed photons for the IC
scattering in this region are not the synchrotron photons, but rather
photons originating from the cosmic microwave background (CMB) or
extra-galactic background light (EBL). The incoming flux of the seed
photons is therefore $(\nu F_\nu)_{\rm peak, in} = \pi r^2 c u_{\rm
  ex}/4 \pi d_L^2$, where $r=R_{\rm acc}$ is the characteristic size
of the acceleration region, and $u_{\rm ex}$ is the energy density of
the CMB/EBL radiation fields. Substituting this result in the
definition of $\bar R$ in equation (\ref{eq:4}), one obtains the
minimum radius which allows acceleration of UHECRs,
\beq
\ba{lcl}
R_{\rm acc} & \geq & \left(\frac{(\nu  F_\nu)_{\rm peak, IC}}{(\nu  F_\nu)_{\rm
      peak, syn}}\right)^{1/2} \left(\frac{1}{8 \pi
    u_{\rm ex}}\right)^{1/2} \left(\frac{\eta E^{ob}}{Z q}\right) \beta
\mathcal{D} \\
& = & 35 \left(\frac{(\nu  F_\nu)_{\rm peak, IC}}{(\nu  F_\nu)_{\rm
      peak, syn}}\right)^{1/2}  \left(\frac{\eta E_{20}^{ob}}{Z}\right) \beta \mathcal{D} 
\mathcal  {\rm \; kpc}.
\ea
\label{eq:6}
\eeq
In evaluating the minimum radius in the second line of equation
(\ref{eq:6}), we conservatively used $u_{CMB} \simeq 4 \times 10^{-13}
{\rm \; erg \, cm^{-3}}$. While several models exist for the energy
density of the EBL light \cite{Georg+08}, it is generally smaller,
$u_{EBL} \sim 10^{-14} {\rm \; erg \, cm^{-3}}$. Thus, if EBL photons
are the target photons, a tighter constraint on the acceleration radius
is obtained. The allowed region is plotted in Figure
\ref{fig:2}.\footnote{Note that the origin of the inequality lies in
  the Hillas condition, equation (1).}

\paragraph*{Acceleration of UHECRs in Cen A.} At distance
of $\approx 3.7$~Mpc \cite{Ferrarese+07}, Cen~A is the nearest
AGN. Due to its proximity, it was long been considered as a possible
source of UHECRs \cite{Cavallo78,Romero+96}. A renewed interest in its
ability to produce UHECRs was prompted recently by the realization
that the arrival directions of two and potentially even four UHECRs
coincide within errors with its position \cite{Abraham+07, GTTT08,
  Moskalenko+09}. By analyzing VLA data, \cite{Hardcastle+03} concluded
that the inner jet is trans-relativistic, $\beta \sim 0.5$, with
Doppler shift $\mathcal{D} \gtrsim 1$. A similar conclusion was
recently drawn by the Fermi team \cite{Abdo+10}, where $\mathcal{D}
\sim {\rm few}$ was estimated, based on broad-band fits to the
spectrum.   

Emission from Cen~A's giant radio lobes, which extend to $\gtrsim
300$~kpc, was recently resolved by the {\it Wilkinson Microwave
  Anisotropy Probe (WMAP)} \cite{Hardcastle+09}. By analyzing the data
and estimating the magnetic field strength and the jet magnetic
luminosity, it was suggested by Ref. \cite{Hardcastle+09} that the
giant lobes are possible acceleration sites of UHECRs.  The giant
lobes were later resolved by the {\it Fermi} observatory
\cite{Abdo+10b}, constraining $(\nu F_\nu)_{\rm peak, IC} /(\nu
F_\nu)_{\rm peak, syn} \geq 10$.

The results presented in Figures \ref{fig:1} and \ref{fig:2}, which
are based on broad band data, show that the core region of Cen~A does
not fulfill the requirements that would enable it to accelerate
protons to $10^{20}$~eV. However, the results presented in equation
(\ref{eq:6}) and Figure \ref{fig:2} do not rule out the proposal
\cite{Hardcastle+09} that acceleration to $10^{20}$~eV could occur in
the outer regions of the giant lobes.  Using the additional data
enabled by {\it Fermi}, we add a constraint on the minimum
acceleration radius to be $\gtrsim 110$~kpc.

An upper limit on the flux of UHECRs from Cen~A is obtained by
assuming that all 13 events seen within $18^\circ$ of
Cen~A \cite{Abreu+10} indeed originate from its lobes. This implies an
observed cosmic rays flux $\mathcal{J} \approx 2 \times 10^{-17} {\rm
  \, m^{-2} s^{-1} sr^{-1}}$, which is translated into luminosity $\sim
4 \pi d_L^2 \mathcal{J} E^{ob} = 10^{38.5} {\rm \, erg \, s^{-1}}$. This
value is about three orders of magnitude less than the synchrotron
luminosity of Cen~A (see Figure \ref{fig:1}), implying energetic
  consistency, even when interpolating the UHECRs flux to lower
  energies, provided that the power law index is not significantly
  greater than 2.

\paragraph*{Discussion.} We derived constraints on the ability of
AGNs to accelerate UHECRs. These constraints are set as limits on
observable quantities, in particular the synchrotron peak luminosity
and the ratio of IC to synchrotron peaks. We showed in equation
(\ref{eq:5}) and Figure \ref{fig:1} that the inner core regions of Cen~A
and M87 do not fulfill the necessary requirement, and therefore cannot
be sources of UHECRs at $10^{20}$~eV, unless UHECRs are composed of
heavy nuclei.

On the other hand, we showed that outer regions of giant radio lobes,
which are frequently seen in radio galaxies, are possible acceleration
sites. Using {\it WMAP} and {\it Fermi} data, we derived in equation
(\ref{eq:6}) and Figure \ref{fig:2} the constraint that the
acceleration site of the closest radio galaxy, Cen~A, must be $\gtrsim
110$~kpc.

A common criticism of the possibility that AGNs might be the main
sources of UHECRs is the lack of sources with bolometric luminosity
$L_{\rm bol} > 10^{46} \llu$ within the GZK horizon \cite[e.g.,][and
references therein]{Lemoine09, Waxman11}, requiring potentially bright
flares \cite{FG09} \cite[but see][]{WL09}.  However, the origin of this
requirement lies in the assumption of {\it equipartition} of energy
between photons and magnetic field, and is therefore not of general
validity.

Indeed, due to the difficulty in measuring the magnetic field strength,
equipartition (or close to it) is often assumed \cite{PS92, Norman+95,
  Levinson06, LO07}. Based on this assumption it was claimed by \cite{PS92}
that AGN cores are plausible acceleration sites. However, this claim
did not consider the limitation of the observed luminosity in nearby
AGNs. Moreover, as was shown in \cite{Norman+95}, energetic protons
cannot escape inner regions without significantly energy loss due to
photo-pion production, which limits the maximum energy of observed
UHECRs.  It was further pointed out by Ref. \cite{Henri+99} that
acceleration of CRs to high energies during flaring activities is
limited due to insufficient residence time in the accelerated region.

We presented a refined analysis, based on identification of the
synchrotron and the IC peaks in broad band spectra of AGNs, which
became available for Cen~A and M87 through observations by the {\it
  Swift} and {\it Fermi} observatories. By doing so, we removed any
dependence on the limited validity of the equipartition
assumption. This analysis applies to 'dark' sources, namely those
sources in which the energy density in the magnetic field is much
larger than the energy density in the photon field. This is indeed the
case for the giant radio lobes of Cen~A, where at $\sim 100$~kpc, the
data shows that the ratio of magnetic to photon energy densities is
$\sim 10^6$ \cite{Hardcastle+09}.

A detailed spectral analysis of many blazars was carried by
\cite{CG08}. In this analysis, a leptonic model (synchrotron and
synchrotron self-Compton) was used to fit the broad band spectra of 73
nearby blazars (all of which are outside the GZK horizon). The
obtained fits thus do not rely on the equipartition assumption. This
analysis is therefore very similar to the one presented here, even
more detailed. The fact that many blazars were found to have similar
spectral shapes thus hints towards a possible generalization of our
key results to many nearby AGNs.

Using values obtained in a similar fitting to the broad-band spectra
in Cen~A \cite{CCC01}, it was claimed by \cite{LW09} that the core
region of Cen~A is excluded as an acceleration site of UHECRs. This
conclusion is similar to the conclusion derived here, based on a much
newer and better quality {\it Fermi} data. The analysis carried by
Ref. \cite{LW09}, however,
did not consider the external giant radio lobes.

In the analysis presented here we use a simplified assumption, namely
that the electrons are mono-energetic. Thus, we ignore the energy
distribution, which is needed for full spectral fits. Nonetheless,
including a broad band spectral fitting, as was done, e.g., in Ref.
\cite{Abdo+10} does not change any of our conclusions. This is because
only a small fraction of the electrons are being accelerated to high
energies, and most of the emission occurs near the electron energy
responsible for the synchrotron peak.  This was verified by comparing
numerical modeling of GRB spectra to simplified analytical
approximations \cite{PW04}.

While our analysis excluded the core regions of Cen~A and M87 (and
plausibly other nearby AGNs) from being sources of UHECRs, giant
flares may change this conclusion.  However, no measurement of the
Compton parameter during flares exists. If such a measurement becomes
available, it could further constrain the ability of AGN cores to
produce UHECRs during flaring activities.

While existing data excludes the cores of Cen~A and M87 from being
potential acceleration sites of protons to $10^{20}$~eV, this is not
the case if UHECRs are composed of heavy nuclei. Recent
data collected by the {\it Pierre Auger Observatory (PAO)} show clues
for heavy composition \cite{Abraham+10}. This data, however,  is in
contrast to the results obtained by the HiRes detector
\cite{Abbasi+10}, and is therefore tentative. We point out
though, that if indeed UHECRs are composed of heavy nuclei, not only
the cores of radio loud AGNs are potential acceleration sites, but so
many other objects, such as radio quiet AGNs \cite{PMM09}.

\acknowledgments
We would like to thank Shmulik Balberg, Ramesh Narayan, Uri Keshet,
Peter M\'esz\'aros and Colin Norman for useful
discussions.  This work was supported in part by NSF grant AST-0907890
and NASA grants NNX08AL43G and NNA09DB30A.

\bibliographystyle{/Users/apeer/Documents/Bib/apsrev}


\end{document}